\documentclass[useAMS]{article}

\usepackage{graphicx}
\begin{document}
\title{{\itshape Two Algorithms for Solving A General Backward Pentadiagonal Linear Systems}}

\author{ A. A.KARAWIA\footnote{ Home Address: Mansoura University,
Mansoura 35516, Egypt. e-mail: abibka@mans.edu.eg}\\
\\
Department of Mathematics, Faculty of Science, Qassim University, \\
  Buraidah, 51452, KSA}

\date{}

\makeatletter \@addtoreset{equation}{section}
\renewcommand{\theequation}{\thesection.\@arabic\c@equation}
\makeatother

\maketitle
\begin{abstract}
In this paper we present an efficient computational and symbolic
algorithms for solving  a backward pentadiagonal linear systems. The
implementation of the algorithms using Computer Algebra Systems
(CAS) such as MAPLE, MACSYMA, MATHEMATICA, and MATLAB are
straightforward. An examples are given in order to illustrate the
algorithms. The symbolic algorithm is competitive the
 other methods for solving a backward pentadiagonal linear systems.\bigskip

\begin{flushleft}\footnotesize
{\textbf{Keywords:} backward pentadiagonal matrices; pentadiagonal
matrices; linear systems; determinants; computer algebra systems
(CAS).\bigskip

}
\end{flushleft}

\newtheorem{alg}{Algorithm}[section]

\end{abstract}

\section{Introduction}

Many problems in mathematics and applied science require the
solution of linear systems having a backward pentadiagonal
coefficient matrices. This kind of linear system arise in many
fields of
numerical computation [1, 2]. This article is a general case of the author article [3].\\

The main goal of the current paper is to develop an efficient
algorithms for solving a general backward pentadiagonal linear
systems of the form:
\begin{equation}
AX=Y
\end{equation}
where
\begin{eqnarray}
A=\left[\begin{array}{ccccccc}
          0 & 0 & 0 & \cdots & \tilde{a}_1 & a_1 & d_1 \\
          0 & 0 & \cdots & \tilde{a}_2& a_2 & d_2 & b_2 \\
          0 & \cdots& \tilde{a}_3& a_3 & d_3 & b_3 & \tilde{b}_3 \\
          \vdots & \vdots & \cdots & \cdots & \cdots & \cdots & \vdots \\
          \tilde{a}_{n-2}& a_{n-2} & d_{n-2} & b_{n-2}& \tilde{b}_{n-2} & \cdots & 0 \\
          a_{n-1} & d_{n-1} & b_{n-1} & \tilde{b}_{n-1}& \cdots & 0 & 0 \\
          d_n & b_n & \tilde{b}_n & \cdots & 0 & 0 & 0
        \end{array}\right],
\end{eqnarray}
$X=(x_1,x_2,\ldots,x_n)^T$, $Y=(y_1,y_2,\ldots,y_n)^T$ and $n\ge5$.\\
\\
A general $n\times n$ backward pentadiagonal matrix $A$ of the form
(1.2) can be stored in $5n-6$ memory locations by using five vectors
$\mbox{\boldmath$\tilde{a}$}=(\tilde{a}_1,\tilde{a}_2,\ldots,\tilde{a}_{n-2})$,
$\mbox{\boldmath$a$}=(a_1,a_2,\ldots,a_{n-1})$,
$\mbox{\boldmath$b$}=(b_2,b_3,\ldots,b_n)$,
$\mbox{\boldmath$\tilde{b}$}=(\tilde{b}_3,\tilde{b}_4,\ldots,\tilde{b}_n)$,
and $\mbox{\boldmath$d$}=(d_1,d_2,\ldots,d_n)$. When considering the
system (1.1) it is advantageous to introduce three additional
vectors
$\mbox{\boldmath$\alpha$}=(\alpha_1,\alpha_2,\ldots,\alpha_{n-1})$,
$\mbox{\boldmath$\beta$}=(\beta_1,\beta_2,\ldots,\beta_n)$ and
$\mbox{\boldmath$\gamma$}=(\gamma_2,\gamma_3,\ldots,\gamma_n)$. These
vectors are related to the vectors $\mbox{\boldmath$\tilde{a}$}$,
$\mbox{\boldmath$a$}$, $\mbox{\boldmath$d$}$, $\mbox{\boldmath$b$}$, and $\mbox{\boldmath$\tilde{b}$}$.\\

 The current paper is organized as follows. In section 2, the main
 results are given. An illustrative examples are presented in section 3. In section 4, a
 Conclusion is given.

\section{Main results}
In this section we are going to formulate a modified computational
algorithm and an efficient symbolic algorithm for solving  a general
backward pentadiagonal linear systems of the form (1.1). To do this
we begin by translate the system (1.1) to the following
pentadiagonal linear system:
\begin{equation}
A_1X=Y_1
\end{equation}
where
\begin{eqnarray}
A_1=\left[\begin{array}{ccccccc}
          d_n & b_n & \tilde{b}_n & \cdots & 0 & 0 & 0 \\
          a_{n-1} & d_{n-1} & b_{n-1}& \tilde{b}_{n-1} & \cdots & 0 & 0 \\
          \tilde{a}_{n-2} & a_{n-2} & d_{n-2} & b_{n-2}& \tilde{b}_{n-2} & \cdots & 0 \\
          \vdots & \ddots & \ddots & \ddots & \ddots & \ddots & \vdots \\
          0 & \cdots &\tilde{a}_3& a_3 & d_3 & b_3 & \tilde{b}_3 \\
          0 & 0 & \cdots &\tilde{a}_2& a_2 & d_2 & b_2 \\
          0 & 0 & 0 & \cdots & \tilde{a}_1 & a_1 & d_1
        \end{array}\right],
\end{eqnarray}
$Y_1=(y_n,y_{n-1},\ldots,y_1)^T$ and $n\ge5$.\\
\\
Now considering the $LU$ decomposition [4] of the matrix $A_1$ in the form:\\
\begin{eqnarray}
\left[\begin{array}{ccccccc}
          d_n & b_n & \tilde{b}_n & \cdots & 0 & 0 & 0 \\
          a_{n-1} & d_{n-1} & b_{n-1}& \tilde{b}_{n-1} & \cdots & 0 & 0 \\
          \tilde{a}_{n-2} & a_{n-2} & d_{n-2} & b_{n-2}& \tilde{b}_{n-2} & \cdots & 0 \\
          \vdots & \ddots & \ddots & \ddots & \ddots & \ddots & \vdots \\
          0 & \cdots &\tilde{a}_3& a_3 & d_3 & b_3 & \tilde{b}_3 \\
          0 & 0 & \cdots &\tilde{a}_2& a_2 & d_2 & b_2 \\
          0 & 0 & 0 & \cdots & \tilde{a}_1 & a_1 & d_1
        \end{array}\right]=\nonumber
\end{eqnarray}
\begin{eqnarray}
          \left[\begin{array}{ccccccc}
          1 & 0 & 0 & \cdots & 0 & 0 & 0 \\
          \gamma_2 & 1 & 0 & \cdots & 0 & 0 & 0 \\
          \frac{\tilde{a}_{n-2}}{\beta_1} & \gamma_3 & 1 & 0 & \cdots & 0 & 0 \\
          \vdots & \ddots & \ddots & \ddots & \ddots & \cdots & \vdots \\
          0 & \cdots &\frac{\tilde{a}_3}{\beta_{n-4}}& \gamma_{n-2} & 1 & 0 & 0 \\
          0 & 0 & \cdots & \frac{\tilde{a}_2}{\beta_{n-3}}& \gamma_{n-1} & 1 & 0 \\
          0 & 0 & \cdots & 0 & \frac{\tilde{a}_1}{\beta_{n-2}}& \gamma_n & 1
          \end{array}\right]\left[\begin{array}{ccccccc}
          \beta_1 & \alpha_1 & \tilde{b}_n & 0 & \cdots & 0 & 0 \\
          0 & \beta_2 &  \alpha_2 & \tilde{b}_{n-1} & \cdots & 0 & 0 \\
          0 & 0 & \beta_3 &  \alpha_3 & \tilde{b}_{n-2} & \cdots & 0 \\
          \vdots & \vdots & \cdots & \ddots & \ddots & \ddots & \vdots \\
          0 & 0 & 0 & \cdots & \beta_{n-2} & \alpha_{n-2} &\tilde{ b}_3\\
          0 & 0 & 0 & \cdots & 0 & \beta_{n-1} & \alpha_{n-1} \\
          0 & 0 & 0 & \cdots & 0 & 0 & \beta_n
        \end{array}\right].
\end{eqnarray}
From (2.3) we obtain
\begin{equation}
\alpha_i= \left\{\begin{array}{ll}
b_n & \textrm{if $i=1$}\\
b_{n-i+1}-\gamma_i \tilde{b}_{n-i+2} & \textrm{if $i=2(1)n-1$,}
\end{array}\right.
\end{equation}
\\
\begin{equation}
\beta_i= \left\{\begin{array}{ll}
d_n & \textrm{if $i=1$}\\
d_{n-1}-\alpha_1\gamma_2 & \textrm{if $i=2$}\\
d_{n-i+1}-\frac{\tilde{a}_{n-i+1}}{\beta_{i-2}}\tilde{b}_{n-i+3}-\alpha_{i-1}\gamma_i
& \textrm{if $i=3(1)n$,}
\end{array}\right.
\end{equation}
\\
and
\begin{equation}
\hspace*{-2.7cm}\gamma_i= \left\{\begin{array}{ll}
\frac{a_{n-1}}{\beta_1} & \textrm{if $i=2$}\\
\frac{a_{n-i+1}-\frac{\tilde{{a}}_{n-i+1}}{\beta_{i-2}}\alpha_{i-2}}{\beta_{i-1}}
& \textrm{if $i=3(1)n$.}
\end{array}\right.
\end{equation}
\\
It is not difficult to prove that the $LU$ decomposition (2.3)
exists only if $\beta_i\ne 0,\quad i=1(1)n-1$. Moreover a general
backward pentadiagonal linear system (1.1) possesses a unique
solution if, in addition, $\beta_n\ne 0$. On the other hand, the
Determinant of the matrix $A_1$ is given by
\begin{equation}
det(A_1)=\prod_{i=1}^n \beta_i,
\end{equation}
and this shown the importance of the vector
$\mbox{\boldmath$\beta$}$ [5].\\
\\
We may now formulate the following results.
\begin{alg}
To solve the general backward pentadiagonal linear system (1.1),
we may procced as follows:\\
\textbf{step 1:} Set $\beta_1=d_n$, $\gamma_2=\frac{a_{n-1}}{\beta_1}$, $\alpha_1=b_n$,
 $\beta_2=d_{n-1}-\alpha_1*\gamma_2$, and $\alpha_2=b_{n-1}-\gamma_2*\tilde{b}_n$.\\
\textbf{step 2:} If  $\beta_1=0$ Or $\beta_2=0$, then OUTPUT('the
method is fails'); STOP.\\
\textbf{step 3:} For $i=3, 4,\ldots,n-1$ Compute\\
\hspace*{2.3cm}$\gamma_i=\frac{a_{n-i+1}-\frac{\tilde{a}_{n-i+1}}{\beta_{i-2}}\alpha_{i-2}}
{\beta_{i-1}}$,\\
\hspace*{2.3cm}$\alpha_i=b_{n-i+1}-\gamma_i \tilde{b}_{n-i+2}$,\\
\hspace*{2.3cm}$\beta_i=d_{n-i+1}-\frac{\tilde{a}_{n-i+1}}{\beta_{i-2}}\tilde{b}_{n-i+3}
- \alpha_{i-1} \gamma_i$,\\
\hspace*{2.3cm}If  $\beta_i=0$, then OUTPUT('the method is
fails'); STOP.\\
\textbf{step 4:} Compute\\
\hspace*{2.3cm}$\gamma_n=\frac{a_1-\frac{\tilde{a}_1}{\beta_{n-2}}\alpha_{n-2}}
{\beta_{n-1}}$,\\
\hspace*{2.3cm}$\beta_n=d_1-\frac{\tilde{a}_1}{\beta_{n-2}}\tilde{b}_3
- \alpha_{n-1} \gamma_n$.\\
\textbf{step 5:} Set $z_1=y_n$, $z_2=y_{n-1}-\gamma_2 z_1$.\\
\textbf{step 6:} For $i=3,4, \ldots,n$ Compute\\
\hspace*{2.3cm}$z_i=y_{n-i+1}-\frac{\tilde{a}_{n-i+1}}{\beta_{i-2}}z_{i-2}-\gamma_i
z_{i-1}$.\\
\textbf{step 7:} Compute the solution vector
$\mbox{\boldmath$x$}$ using\\
\hspace*{2cm} $x_n=\frac{z_n}{\beta_n}$,
$x_{n-1}=\frac{z_{n-1}-\alpha_{n-1} x_n}{\beta_{n-1}}$,\\
\hspace*{2cm}For $i=n-2,n-3,\ldots, 1$
compute\\
\hspace*{2.9cm}$x_i=\frac{z_i-\alpha_i x_{i+1}-\tilde{b}_{n-i+1}x_{i+2}}{\beta_{i}}$.\\
\end{alg}
The new algorithm 2.1 will be referred to as \textbf{KBPENTA}
algorithm. \textbf{KBPENTA} algorithm for solving the backward
pentadiagonal system (1.1) is generally preferable because the
conditions $\beta_i\ne 0,\quad i=1(1)n$ are sufficient for the
validity of it. The advantage of the vector
$\mbox{\boldmath$\beta$}$ is now
clear.\\

The following symbolic algorithm is developed in order to remove the
cases where the numeric algorithm \textbf{KBPENTA} fails.

\begin{alg}
To solve the general backward pentadiagonal linear system (1.1),
we may procced as follows:\\
\textbf{step 1:} Set $\beta_1=d_n$.\\
\textbf{step 2:} Set  $\beta_1=x$($x$ is just a symbolic name) whenever $\beta_1=0$.\\
\textbf{step 3:} Set $\gamma_2=\frac{a_{n-1}}{\beta_1}$,
$\alpha_1=b_n$,
 $\beta_2=d_{n-1}-\alpha_1\gamma_2$, and $\alpha_2=b_{n-1}-\gamma_2\tilde{b}_n$.\\
\textbf{step 4:} Set  $\beta_2=x$ whenever $\beta_2=0$.\\
\textbf{step 5:} For $i=3,4,\ldots,n-1$ Compute\\
\hspace*{2.3cm}$\gamma_i=\frac{a_{n-i+1}-\frac{\tilde{a}_{n-i+1}}{\beta_{i-2}}\alpha_{i-2}}
{\beta_{i-1}}$,\\
\hspace*{2.3cm}$\alpha_i=b_{n-i+1}-\gamma_i \tilde{b}_{n-i+2}$,\\
\hspace*{2.3cm}$\beta_i=d_{n-i+1}-\frac{\tilde{a}_{n-i+1}}{\beta_{i-2}}\tilde{b}_{n-i+3}
- \alpha_{i-1} \gamma_i$,\\
\hspace*{2.3cm} Set  $\beta_i=x$ whenever $\beta_i=0$.\\
\textbf{step 6:} Compute\\
\hspace*{2.3cm}$\gamma_n=\frac{a_1-\frac{\tilde{a}_1}{\beta_{n-2}}\alpha_{n-2}}
{\beta_{n-1}}$,\\
\hspace*{2.3cm}$\beta_n=d_1-\frac{\tilde{a}_1}{\beta_{n-2}}\tilde{b}_3
- \alpha_{n-1} \gamma_n$.\\
\textbf{step 7:} Set  $\beta_n=x$ whenever $\beta_n=0$.\\
\textbf{step 8:} Set $z_1=y_n$, $z_2=y_{n-1}-\gamma_2 z_1$.\\
\textbf{step 9:} For $i=3,4, \ldots,n$ Compute\\
\hspace*{2.3cm}$z_i=y_{n-i+1}-\frac{\tilde{a}_{n-i+1}}{\beta_{i-2}}z_{i-2}-\gamma_i
z_{i-1}$.\\
\textbf{step 10:} Compute the solution vector
$\mbox{\boldmath$x$}$ using\\
\hspace*{2cm} $x_n=\frac{z_n}{\beta_n}$,
$x_{n-1}=\frac{z_{n-1}-\alpha_{n-1} x_n}{\beta_{n-1}}$,\\
\hspace*{2cm}For $i=n-2,n-3,\ldots, 1$
compute\\
\hspace*{2.9cm}$x_i=\frac{z_i-\alpha_i x_{i+1}-\tilde{b}_{n-i+1}x_{i+2}}{\beta_{i}}$.\\
\textbf{step 11:} Substitute $x=0$ in all expressions of the
solution vector $x_i,i=1,2,\ldots,n$.\\
\end{alg}
The symbolic algorithm 2.2 will be referred to as \textbf{KSBPENTA}
algorithm.\\
In [6], Claerbout showed that the two-dimensional Laplacian
operator, which appears in 3-D finite-difference migration, has the
form of pentadiagonal matrix. If we choice $d_i=-4,\quad i=1(1)n$,
$a_i=b_i=\tilde{a}_i=\tilde{b}_i=1$, and $y_i=y_j,\quad i=1(1)n,
j=n(-1)1$, we can obtain it.

\section{An Illustrative Examples}
In this section we are going to give an illustrative examples\\
\\
\textbf{Example 3.1.} Solve the backward pentadiagonal linear system
of size $5$ given by
\begin{eqnarray}
\left[\begin{array}{ccccc}
  0 & 0 & 3 & -1 & 1 \\
  0 & 2 & -2 & 2 & 4 \\
  3 & 1 & 2 & 1 & 1 \\
  4 & -2 & 2 & 2 & 0 \\
  -1 & 1 & 1 & 0 & 0
\end{array}\right]\mbox{\boldmath$X$}=\left[\begin{array}{c}
                   10\\
                   26\\
                   20\\
                   14\\
                   4\\
                   \end{array}\right]
\end{eqnarray}
by using the \textbf{KBPENTA} algorithm and \textbf{KSBPENTA} algorithm.\\
\\
\textbf{Solution}\\

(i) The application of the \textbf{KBPENTA} algorithm gives:\\
\begin{itemize}
    \item $\beta_1=-1$, $\gamma_2=-4$, $\alpha_1=1$,$\beta_2=2$, and $\alpha_2=6$ (Step 1).\\
    \item $[\gamma_3, \gamma_4]=[2,\frac{8}{7},-\frac{2}{3}]$,
    $[\beta_3,\beta_4]=[-7,\frac{24}{7}]$, $[\alpha_3, \alpha_4]=
    [-3,\frac{20}{7}]$ (Step 3).\\
    \item $\gamma_5=-\frac{2}{3}$, $\beta_5=\frac{10}{3}$ (Step 4).\\
    \item $z_1=4$, and $z_2=30$ (Step 5).\\
    \item $[z_3,z_4,z_5]=[-28,28,\frac{50}{3}]$ (Step 6).\\
    \item $x_5=5, x_4=4, [x_1,x_2,x_3]=[1, 2, 3]$ (Step 7).
    \end{itemize}
Also the determinant of the matrix $A_1$ is
$det(A_1)=160$ by using (2.7).\\
\\
(ii) The application of the \textbf{KSBPENTA} algorithm gives:\\
\hspace*{2.3cm}$X:=back\_penta(\tilde{b},b,d,a,\tilde{a},y)=[1, 2, 3, 4, 5]$.\\
\\
\textbf{Example 3.2.} Solve the backward pentadiagonal linear system
of size $5$ given by
\begin{eqnarray}
\left[\begin{array}{ccccc}
  0 & 0 & 3 & -1 & 1 \\
  0 & 2 & -2 & 2 & 4 \\
  3 & 1 & 2 & 1 & 1 \\
  4 & -2 & 2 & 2 & 0 \\
  0 & 1 & 1 & 0 & 0
\end{array}\right]\mbox{\boldmath$X$}=\left[\begin{array}{c}
                   10\\
                   26\\
                   20\\
                   14\\
                   5\\
                   \end{array}\right]
\end{eqnarray}
by using the \textbf{KBPENTA} algorithm and \textbf{KSBPENTA} algorithm.\\
\\
\textbf{Solution}\\

(i) The application of the \textbf{KBPENTA} algorithm gives:\\
\hspace*{2.3cm}The method is break down since $\beta_1=d_5=0$.\\
\\
\hspace*{0.5cm}(ii) The application of the \textbf{KSBPENTA} algorithm gives:\\
\hspace*{2.3cm}$X:=back\_penta(\tilde{b},b,d,a,\tilde{a},y)$=$[-\frac{11}{9x-11},
 \frac{22(x-1)}{9x-11}, \frac{34x-33}{9x-11}, \frac{51x-44}{9x-11},
 \frac{39x-55}{9x-11}]_{x=0}$\\
\\
\hspace*{7.75cm}=[1, 2, 3, 4, 5].\\
Also the determinant of the matrix $A_1$ is $det(A_1)=88$ and for
more details about how to call this procedure, see appendix 1.

\section{Conclusion}
 The methods described here are a very effective, provided optimal
 LU factorization is used. Our symbolic algorithm is competitive the
 other methods for solving a backward pentadiagonal linear systems
 which appear in many applications.\\
\section{Acknowledgement}

I should like to thank Prof. Dr. M. E. A. El-Mikkawy for several
comments and suggestions.\\
\\
\\
\textbf{Appendix 1. } A Maple procedure for solving a backward pentadiagonal linear systems\\
\\
\textbf{Notes:} The procedure is based on the following results:\\

\hspace*{1.5cm}$x_i=\frac{z_i-\alpha_ix_{i+1}-\tilde{b}_{n-i+1}x_{i+2}}{\beta_{i}}$,
$i=n-2,n-3,\ldots, 1, x_n=\frac{z_n}{\beta_n}, x_{n-1}=\frac{z_{n-1}-\alpha_{n-1}x_n}{\beta_{n-1}}$.\\
where\\
\hspace*{1.9cm}$z_i=y_{n-i+1}-\gamma_iz_{i-1}-\frac{\tilde{a}_{n-i+1}}{\beta_{i-2}}z_{i-2}
$, $i=3, 4, \ldots,n, z_1=y_n, z_2=y_{n-1}-\gamma_2z_1$,\\
\hspace*{1.8cm}$\gamma_i=\frac{a_{n-i+1}-\frac{\tilde{{a}}_{n-i+1}}{\beta_{i-2}}\alpha_{i-2}}{\beta_{i-1}}$,
$i=3, 4,\ldots,n, \gamma_2=\frac{a_{n-1}}{\beta_1}$,\\
\hspace*{1.8cm}$\beta_i=d_{n-i+1}-\frac{\tilde{a}_{n-i+1}}{\beta_{i-2}}\tilde{b}_{n-i+3}-\alpha_{i-1}\gamma_i$,
$i=3, 4,\ldots,n, \beta_1=d_n, \beta_2=d_{n-1}-\alpha_1\gamma_2$,\\
and\\
\hspace*{1.8cm}$\alpha_i=b_{n-i+1}-\gamma_i \tilde{b}_{n-i+2}$,
$i=2, 3, \ldots, n-1, \alpha_1=b_n$.
\newpage
\noindent $> \#$ A Maple Procedure.\\
$> \#$ Written by Dr. A. A. Karawia.\\
$> \#$ To compute the solution of A general backward pentadiagonal
Linear systems.\\
$>$ restart:\\
$>$ with(linalg,vector,vectdim):\\
$>$ back$\_$penta:=proc(bb::vector,b::vector,
d::vector,a::vector,aa::vector,y::vector)\\
 local i,j,k,n; global T,alpha,beta,g,z,X;\\
    n:=vectdim(d):alpha:=array(1..n-1):beta:=array(1..n):
    g:=array(1..n):\\
    z:=array(1..n):X:=array(1..n):\\
   $\#$components of the vectors alpha,beta, and gamma $\#$\\
   beta[1]:=d[n]:\\
   if beta[1]=0 then beta[1]:=x;d[n]:=x;fi:\\
   g[2]:=a[n-1]/beta[1]:alpha[1]:=b[n]:
   beta[2]:=simplify(d[n-1]-alpha[1]*g[2]):\\
   alpha[2]:=simplify(b[n-1]-g[2]*bb[n]):\\
   if beta[2]=0 then beta[2]:=x;fi:\\
   for i from 3 to n-1 do\\
    \hspace*{1.8cm}g[i]:=simplify((a[n-i+1]-aa[n-i+1]*alpha[i-2]/beta[i-2])/beta[i-1]):\\
    \hspace*{1.8cm}alpha[i]:=simplify(b[n-i+1]-g[i]*bb[n-i+2]):\\
    \hspace*{1.8cm}beta[i]:=simplify(d[n-i+1]-aa[n-i+1]*bb[n-i+3]/beta[i-2]-alpha[i-1]*g[i]):\\
    \hspace*{1.8cm}if beta[i]=0 then beta[i]:=x;fi:\\
   end do:\\
g[n]:=simplify((a[1]-aa[1]*alpha[n-2]/beta[n-2])/beta[n-1]):\\
beta[n]:=simplify(d[1]-aa[1]*bb[3]/beta[n-2]-alpha[n-1]*g[n]):\\
  $\#$ To compute the vector Z $\#$\\
  z[1]:=y[n]:z[2]:=y[n-1]-g[2]*z[1]:i:='i':\\
  for i from 2 to n do\\
  \hspace*{1.8cm}z[i]:=simplify(y[n-i+1]-aa[n-i+1]*z[i-2]/beta[i-2]-g[i]*z[i-1]):\\
  end do:\\
  $\#$ To compute the Solution of the system X. $\#$\\
  X[n]:=z[n]/beta[n]:i:='i':\\
  X[n-1]:=simplify((z[n-1]-alpha[n-1]*X[n])/beta[n-1]):\\
  for i from n-1 by -1 to 1 do\\
  \hspace*{1.8cm}X[i]:=simplify((z[i]-alpha[i]*X[i+1]-bb[n-i+1]*X[i+2])/beta[i]):\\
  end do:\\
  $\#$ To compute the determinant T $\#$\\
  T:=subs(x=0,simplify(product(beta[r],r=1..n)));\\
  eval(X):\\
 end:\\
$> \#$ Call no. 1 for the procedure back$\_$penta. $\#$\\
$>$ x:='x':\\
$>$ aa:=vector([3,-1,7,-2]);\\
\hspace*{3cm} aa := [3, -1, 7, -2]\\
$>$a:=vector([2,5,2,3,-5]);\\
\hspace*{3cm}a := [2, 5, 2, 3, -5]\\
$>$ d:=vector([1,3,3,5,6,14]);\\
\hspace*{3cm} d := [1, 3, 3, 5, 6, 14]\\
$>$ b:=vector([0,2,1,2,2,1]);\\
\hspace*{3cm} b := [0, 2, 1, 2, 2, 1]\\
$>$ bb:=vector([0,0,-5,-7,3,-10]);\\
\hspace*{3cm} bb := [0, 0, -5, -7, 3, -10]\\
$>$ y:=vector([6,9,8,1,6,5]);\\
\hspace*{3cm} y := [6, 9, 8, 1, 6, 5]\\
$>$ X:=back$\_$penta(bb,b,d,a,aa,y);\\
\hspace*{3cm} X := [1, 1, 1, 1, 1, 1]\\
$>$ T;\\
\hspace*{3cm}          -8597\\
$> \#$ End of call no. 1. $\#$\\
\\
$> \#$ Call no. 2 for the procedure back$\_$tri. $\#$\\
$>$ x:='x':\\
$>$ aa:=vector([3,-1,7,-2]);\\
\hspace*{3cm} aa := [3, -1, 7, -2]\\
$>$a:=vector([2,5,2,3,-5]);\\
\hspace*{3cm}a := [2, 5, 2, 3, -5]\\
$>$ d:=vector([1,3,3,5,6,0]);\\
\hspace*{3cm} d := [1, 3, 3, 5, 6, 0]\\
$>$ b:=vector([0,2,1,2,2,1]);\\
\hspace*{3cm} b := [0, 2, 1, 2, 2, 1]\\
$>$ bb:=vector([0,0,-5,-7,3,-10]);\\
\hspace*{3cm} bb := [0, 0, -5, -7, 3, -10]\\
$>$ y:=vector([6,9,8,1,6,-9]);\\
\hspace*{3cm} y := [6, 9, 8, 1, 6, -9]\\
$>$ X:=back$\_$penta(bb,b,d,a,aa,y);\\
\hspace*{3cm} X := $[-\frac{1777}{741x-1777},
-\frac{2x+1777}{741x-1777}, \frac{489x-1777}{741x-1777},
\frac{1160x-1777}{741x-1777}, \frac{574x-1777}{741x-1777},
-\frac{182x+1777}{741x-1777}]$\\
$>$ T;\\
\hspace*{3cm}          1777\\
$>$ x:=0:X:=map(eval,op(X));\\
\hspace*{3cm} X := [1, 1, 1, 1, 1, 1] \\
$> \#$ End of call no. 2. $\#$

\end{document}